\documentclass[aps,prl,reprint,showpacs,superscriptaddress]{revtex4-1}
\usepackage[english]{babel}
\usepackage{amsmath,amssymb,bbm,graphicx,color,comment,txfonts}
\usepackage[bookmarks=true,colorlinks,citecolor=blue,urlcolor=blue]{hyperref}
\usepackage{dsfont}
\usepackage[normalem]{ulem}
\usepackage{physics}
\begin{document}

\date{\today}

\title{Measurement-induced dark state phase transitions in long-ranged fermion systems}

\author{T. M\"uller}
\affiliation{Institut f\"ur Theoretische Physik, Universit\"at zu K\"oln, D-50937 Cologne, Germany}
\author{S. Diehl}
\affiliation{Institut f\"ur Theoretische Physik, Universit\"at zu K\"oln, D-50937 Cologne, Germany}
\author{M. Buchhold}
\affiliation{Institut f\"ur Theoretische Physik, Universit\"at zu K\"oln, D-50937 Cologne, Germany}

\begin{abstract}
We identify an unconventional algebraic scaling phase in the quantum dynamics of free fermions with long range hopping, which are exposed to continuous local density measurements. The unconventional phase is characterized by an algebraic entanglement entropy growth, and by a slow algebraic decay of the density-density correlation function, both with a fractional exponent. It occurs for hopping decay exponents $1< p \lesssim 3/2$ independently of the measurement rate. The algebraic phase gives rise to two critical lines, separating it from a critical phase with logarithmic entanglement growth at small, and an area law phase with constant entanglement entropy at large monitoring rates. A perturbative renormalization group analysis suggests that the transitions to the long-range phase are also unconventional, corresponding to a modified sine-Gordon theory. Comparing exact numerical simulations of the monitored wave functions with analytical predictions from a replica field theory approach yields an excellent quantitative agreement. This confirms the view of a measurement-induced phase transition as a quantum phase transition in the dark state of an effective, non-Hermitian Hamiltonian.
\end{abstract}

\maketitle


Measurement-induced phase transitions provide a new angle for understanding quantum dynamics. The tracking of measurement outcomes along single trajectories contains qualitatively more information than a dynamics where these outcomes are not read, or averaged over. In fact, the individual wave function may undergo phase transitions, which are witnessed by suitable statistical analysis of the trajectory ensemble. The linear (in the state) averaged quantum dynamics represented by a Lindblad master equation converges towards a featureless infinite temperature state, where the extensive configurational entropy generated by all possible measurement outcomes erases all the information on the trajectory wave function. In contrast, state dependent, nonlinear observables, e.g., the entanglement entropy $S$ or connected correlation functions, reveal the non-trivial underlying quantum dynamics.  

\begin{figure}[h]
  \includegraphics[width=1\linewidth]{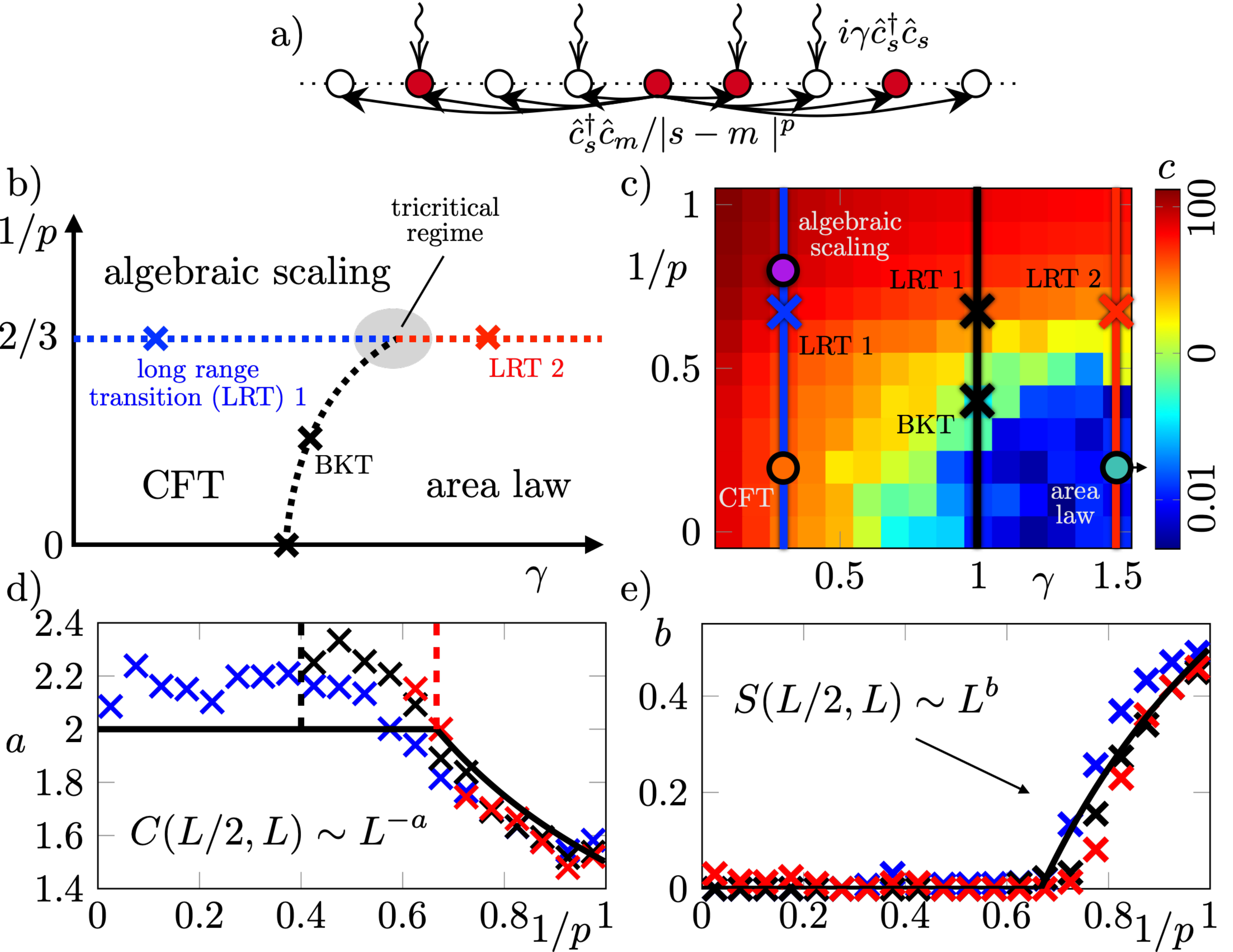}
  \caption{a) Illustration of free fermions on a chain with algebraic  hopping ($1<p<\infty$) and continuous monitoring with rate $\gamma$. b) Schematic phase diagram: we detect and characterize three qualitatively different phases. c) Phase diagram as determined from the effective central charge $c$ for $L=600$. Circles indicate the points discussed in Fig.~\ref{Fig2}. The color coded vertical lines are studied in d,e): comparison of the numerically determined scaling exponents at $\gamma=0.3$ (blue), $\gamma=1$ (black) and $\gamma=1.5$ (red) with the dark state (solid black line). d) Correlation function exponent $a$. Dashed lines yield an estimate for the point where the algebraic scaling terminates 
  and is replaced by exponential scaling. e) Entanglement entropy growth exponent $b$. A vanishing exponent corresponds to logarithmic scaling or area law. The critical points found by this procedure are indicated as crosses in c).}
  \label{Fig1}
\end{figure}

So far, two major types of such transitions have been identified by their entanglement growth with system size $L$. One is represented by transitions from a volume- ($S\sim L$) to an area law ($S\sim L^0$)  \cite{Skinner2019,Fisher2018,Li2019b,gullans2019,choi2020prl,Jian2020,fan2020selforganized,lifisher2021,nahum2021prxq}, and has been established mainly for random circuits, but also for certain interacting Hamiltonians~\cite{fuji2020,jian2021syk,doggen2021generalized}. A second type corresponds to transitions from a critical phase with a logarithmic growth of entanglement entropy ($S\sim \log L$), again to an area law~\cite{alberton2021enttrans, buchhold2021effective, bao2021symmetry,turkeshi2021measurementinduced,ippoliti2020}.

This calls for the question, whether this is the exhaustive set of phases and phase transitions that can exist in monitored quantum dynamics. A promising candidate for exploring it are systems with long-ranged generators of dynamics: Long-ranged Hamiltonians are known to induce new phases in ground states, and to qualitatively modify the critical behavior at phase transitions \cite{campa2014,dutta2001,maghrebi2017}. The thus achieved redistribution of particles over large distances can be expected to enhance the amount of real space entanglement; indeed, entanglement dynamics and correlation spreading are strongly modified by long range contributions to the Hamiltonian \cite{phauke2013,schachenmayer2013,richerme2014}. Experimental platforms for engineering such Hamiltonians range from trapped ions \cite{britton2012,richerme2014,jurcevic2014,zhang2017}, cold atoms in cavities \cite{landig2016,mivehvar2021cavity}, Rydberg atoms \cite{schauss2012}, and polar molecules \cite{yan2013}.

In this work, we explore an elementary model of monitored, long-ranged dynamics: Free fermions with variably ranged hopping, characterized by an algebraic range exponent $p$ (see Fig.~\ref{Fig1}a), competing with disentangling, local particle number measurements. As a main result, we demonstrate that the long-range entangling evolution leads to the emergence of an unconventional dynamical phase, in which the entanglement entropy grows with the system size $S\sim L^{b}$ with $b= 3/2 - p$~\footnote{The numerical results match the analytical predictions with very high accuracy and we use the analytical identities for $a,b,p_c$ here and in the following.}, faster than logarithmically but slower than the volume law of fully ergodic dynamics. Alternatively, the phase can be characterized via nonlinear density-density correlation functions, which follow an algebraic decay $\sim L^{-a}$ with a continuously varying exponent $a= p+1/2$. This phase is realized for exponents $1 < p< p_c = 3/2$, irrespective to the strength of measurement $\gamma >0$: Local  measurements cannot supersede the entanglement generated by long-range coherent hopping. This implies the existence of a tricritical point, where a conformally invariant phase with logarithmic entanglement growth, an area law phase, and the long-range phase meet (see Fig.~\ref{Fig1}b,c). 

To establish these results, we employ a combination of numerical simulations \cite{Cao2019,alberton2021enttrans, turkeshi2021measurementinduced} and an analytical replica field theory \cite{buchhold2021effective}, in which the steady state under monitoring emerges as the dark state of an effective, non-Hermitian sine-Gordon Hamiltonian. In the Hamiltonian framework, measurements induce a pinning of the density fluctuations, and the long-range hopping pins the phase fluctuations. The analytical results align well with the numerical simulations, predicting the critical value of $p$ and the scaling exponents in the long-range phase with high accuracy.
This strengthens the picture of the measurement-induced phase transitions as a quantum phase transition in the dark state of an effective, non-Hermitian measurement-Hamiltonian.

\textit{Microscopic model and effective Hamiltonian.} -- 
We consider fermions on a ring of $L$ sites (labeled $s$) and a dynamics, which is composed of continuous measurements of the local particle number $\hat{n}_s = \hat{c}_s^\dagger \hat{c}_s$ and unitary, long-range hoppings. In the quantum state diffusion protocol, this is described by the stochastic Schrödinger equation (SSE),
\begin{equation} \label{eq:QSD}
    d\ket{\psi \lbrace \xi_{s,t} \rbrace} = \Big[ -i \hat{H} dt + \sum_s \Big( \xi_{s,t} \hat{M}_{s,t}-\tfrac{\gamma}{2} \hat{M}_{s,t}^2 dt \Big) \Big] \ket{\psi \lbrace \xi_{s,t} \rbrace}.
\end{equation}
Here, $\gamma$ is the dimensionless monitoring strength, $\hat{M}_{s,t}=\hat{n}_s-\bra{\psi \lbrace \xi_{s,t} \rbrace} \hat{n}_s \ket{\psi \lbrace \xi_{s,t} \rbrace}$ are the {\it monitoring operators} and $\xi_{s,t}$ is a Gaussian white noise with zero mean $\overline{\xi_{s,t}}=0$ and short-ranged correlations $\overline{\xi_{s,t} \xi_{m,t'}} = \gamma dt \delta_{s,m} \delta(t-t')$. The overbar denotes the noise or trajectory average.
The quantum state diffusion~\cite{deVega2017,Gisin_1992,Strunz1998} is realized by coupling the local observable $\hat{n}_s$ to a continuum of bath degrees of freedom, e.g., pointers~\cite{Schomerus2019,Milburn1993}, which is read out continuously, e.g., via homodyne detection~\cite{Yang2018}. 

The long-range hopping is described by the Hamiltonian
\begin{equation} \label{Eq:LongRangeHamiltonian}
    \hat{H}=-\sum_{s \neq m} t_{s,m}\hat{c}_s^\dagger \hat{c}_m, \ \ \ t_{s,m}=\vert s-m \vert^{-p},
\end{equation}
where $\hat{c}_s^\dagger, \hat{c}_s$ are fermion creation and annihilation operators, $ \lbrace \hat{c}_s,\hat{c}_m^\dagger \rbrace = \delta_{s,m}$. 
The exponent $p$ determines the range over which a single particle is coherently spread, which increases with smaller $p$. In order to ensure a well-defined thermodynamic limit and a bounded fermion dispersion, we consider $p>1$. For $p \rightarrow \infty$, nearest-neighbor hopping is reproduced~\cite{alberton2021enttrans,buchhold2021effective,Cao2019}. The number conserving, quadratic evolution can be described in terms of efficiently simulated Gaussian wave functions. We start from the half-filled Néel state, and evolve the system until stationarity is reached. 

\begin{figure}
  \includegraphics[width=1\linewidth]{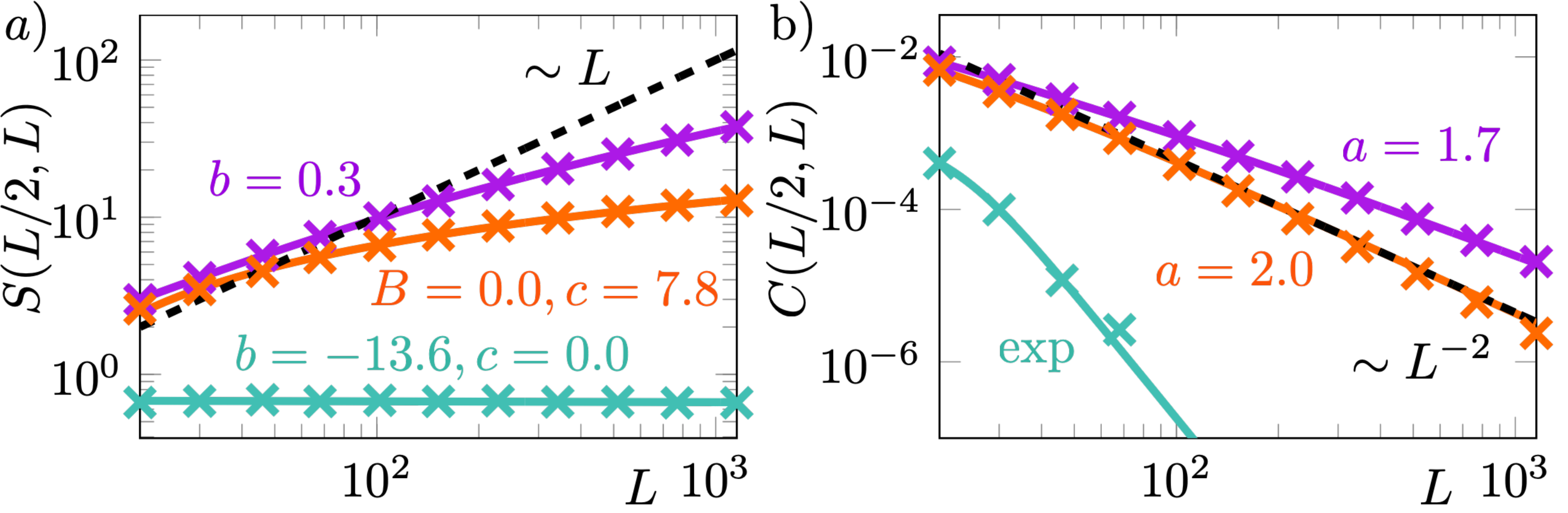}
  \caption{Phase characterization of conformally invariant ($\gamma=0.3, p=1.25$, orange), area-law ($\gamma=2,p=5$, light blue) and algebraic scaling phase ($\gamma=0.3,p=5$, purple). a) Entanglement entropy as a function of system size and best fit using the ansatz $S = \frac{c}{3} \log \frac{L}{\pi} + s_0 + B L^b$. The entropy always grows slower than a volume law ($\sim L$, dashed line). b) The scaling exponent of the correlation function at opposite ends of the system is determined by fitting to $C=1/\left[ A L^a + D L \right]$ (dashed line $\sim L^{-2}$).
  }
  \label{Fig2}
\end{figure}

The probabilistic measurements subject the wave function to a large degree of randomness, which then generates a large configurational entropy: each allowed measurement outcome will appear in the long-time limit with equal probability. As a consequence, the trajectory average of the conditioned density matrix  $\rho_{c,t}=|\psi\{\xi\}\rangle\langle\psi\{\xi\}|$ always yields a maximally mixed state $\overline{\rho_{c,t}}\sim\mathds{1}$. In order to reveal the nontrivial quantum dynamics in each  individual wave function $|\psi\{\xi\}\rangle$, one has to consider trajectory averages of specific observables in which $\rho_{c,t}$ enters nonlinearly~\cite{buchhold2021effective,Nahum2017,nahum2021prxq,Bao_2020}, such as the entanglement entropy or conditioned correlation functions (see below). 

The dynamics of nonlinear observables is determined by a set of {\it replicated} density matrices $\rho_{c,t}\otimes\rho_{c,t}\otimes...$, which are described by a replica field theory~\cite{buchhold2021effective}. It separates the measurement randomness (accumulating it in an effective center-of-mass coordinate) from the inherent quantum evolution of the wave functions, and expresses the latter in terms of an effective, non-Hermitian Hamiltonian. Two-body correlation functions are described by the two-replica product $\rho_{c,t}\otimes\rho_{c,t}$, the replica field theory of which is obtained in three steps: (i) Derive the replica master equation for a product of $2$ replicated density matrices as outlined in Ref.~\cite{buchhold2021effective} (ii) Bosonize the fermion operators on each replica and express the theory in terms of boson field operators $\hat\phi_{x}, \hat\theta_{x}$. Operators on the same replica obey the canonical commutation relation $[\hat\phi_y,\partial_x\hat\theta_x]=i\delta(x-y)$, while they commute for different replicas. (iii) Perform the unitary transformation to the relevant replica coordinates, i.e., the relative and the absolute modes,
    $\hat{\phi}^{(a,r)}_x=\tfrac{1}{\sqrt{2}}(\hat{\phi}^{(1)}_x\pm \hat{\phi}^{(2)}_x),\ \ \hat{\theta}^{(a,r)}_x=\tfrac{1}{\sqrt{2}}(\hat{\theta}^{(1)}_x\pm\hat{\theta}^{(2)}_x)$, where $(1,2)$ label replica $1,2$. 
The absolute modes $\hat{\phi}^{(a)}_x, \hat{\theta}^{(a)}_x$ pick up all the configurational entropy from the replicas ~\cite{buchhold2021effective}. They are structureless and can readily be traced out. 

Executing the steps (i)-(iii) for the SSE in Eq.~\eqref{eq:QSD} reveals that the steady state of the monitoring evolution in the replica framework is a dark state of the effective, non-Hermitian Hamiltonian (dropping the relative replica index)
\begin{subequations}
\begin{align}
    0&=\hat{H}_{\text{eff}}|\psi_D\rangle, \ \ \ \hat{H}_{\text{eff}}=\hat{H}_{\text{sr}}+\hat{H}_{\text{lr}},\label{eq:darkstate}\\
    \hat{H}_{\text{sr}}&=\int_x\left\{ \eta^{-1} (\partial_x \hat{\theta}_x)^2+\eta(\partial_x \hat{\phi}_x)^2-i\lambda\left[1-\cos \sqrt{8}\hat\phi_x \right]\right\}\label{eq:H_sr},\\
    \hat{H}_{\text{lr}}&=i\Delta\int_{x, |y|>1}|y|^{-2p}\left\{1+\cos\left[\sqrt{2}(\hat\theta_x-\hat\theta_{x+y})\right] \right\}.\label{eq:ReplicaHamiltonian}
\end{align}
\end{subequations}
The short-ranged part, $\hat{H}_{\text{sr}}$, covers the nearest neighbor hopping and the measurements of the local particle density. It is of the sine-Gordon form with $\eta^2=1-\tfrac{2\gamma i}{\pi}$ and $\lambda>0$, which has been derived in Ref.~\cite{buchhold2021effective}.

The long-range part, $\hat{H}_{\text{lr}}$ corresponds to hopping across more than one site $\hat{c}^\dagger_l \hat{c}_{l+m}$,  $m>1$. In bosonization $\hat{c}^\dagger_l \hat{c}_{l+m}\sim e^{i(\hat\theta_l-\hat\theta_{l+m})}(1+{\text{h.h.}})$, and we neglect contributions from higher harmonics $\sim$h.h.  (see~\cite{supp} for a detailed discussion). 
The effective Hamiltonian has two remarkable features: Firstly, due to the monitoring $\hat{H}_{\text{lr}}$ is purely imaginary ($\Delta>0$), although the original fermion hopping is Hermitian. Secondly, the effective hopping amplitude decays with an exponent $\sim2p$, which is twice as large as for the microscopic amplitude. Both of these features are emergent in the replica theory and leave clear signatures in the fermion observables.

\textit{Observables.} -- 
In order to characterize the steady state of the SSE \eqref{eq:QSD}, we compute the trajectory averages of the von-Neumann entanglement entropy $S$ and of the connected density-density correlation function $C$ for fixed system size $L$,
\begin{subequations}
\begin{align}
     S(l,L) &= \overline{-\Tr \rho_l \log \rho_l}, \\
     C(l,L) &= \overline{\langle \hat{n}_{x} \hat{n}_{x+l} \rangle - \langle \hat{n}_{x} \rangle \langle \hat{n}_{x+l} \rangle}=\overline{|\langle \hat c^\dagger_{x}\hat c_{x+l} \rangle|^2} .\label{eq:corr_def}
\end{align}
\end{subequations}
Here, $\rho_l=\Tr_{L \setminus l} \ket{\psi \lbrace \xi \rbrace} \bra{\psi \lbrace \xi \rbrace}$ is the reduced density matrix of a subsystem of length $l$ for a pure state $\ket{\psi \lbrace \xi \rbrace}$. In the replica field theory, where we take $L\to \infty$, their leading behavior is captured by a simple form~\cite{supp},
\begin{equation} \label{eq:scanalyyt}
  S(l)= \tfrac{1}{3}\langle\hat\phi_x\hat\phi_{x+l}\rangle, \quad   C(l)=\tfrac{1}{2\pi^2}\partial_l^2\langle\hat\phi_x\hat\phi_{x+l}\rangle.
\end{equation}
Here the expectation value is taken with respect to the dark state $|\psi_D\rangle$ of $\hat{H}_{\text{eff}}$, Eq.~\eqref{eq:darkstate}. The form of $C(l)$ follows from $\hat{n}_x\rightarrow\rho_0-\frac{\partial_x \hat\phi_x}{\pi}$, and the entropy formula is obtained under the assumption of free Dirac fermions in Gaussian states~\cite{Casini_2009, Calabrese_2004}.

\textit{Dark state phase structure.} -- We find three qualitatively different regimes (see Fig.~\ref{Fig1}b): (i) An area law phase for large monitoring ($\gamma \gtrapprox 1$) and short-ranged hopping ($p>3/2$), (ii) a conformally invariant (CFT) phase for smaller but non-vanishing monitoring and short-ranged hopping ($p>3/2$), and (iii) a novel algebraic scaling phase due to long-range hopping ($p<3/2$). Additionally, isolated at $\gamma=0$, a volume law is realized. This obstructs the observation of the large-distance behavior in the dynamical phases for insufficient system sizes~\cite{alberton2021enttrans}. In Fig.~\ref{Fig2}, we show representative results for each phase.

\begin{figure}
  \includegraphics[width=1\linewidth]{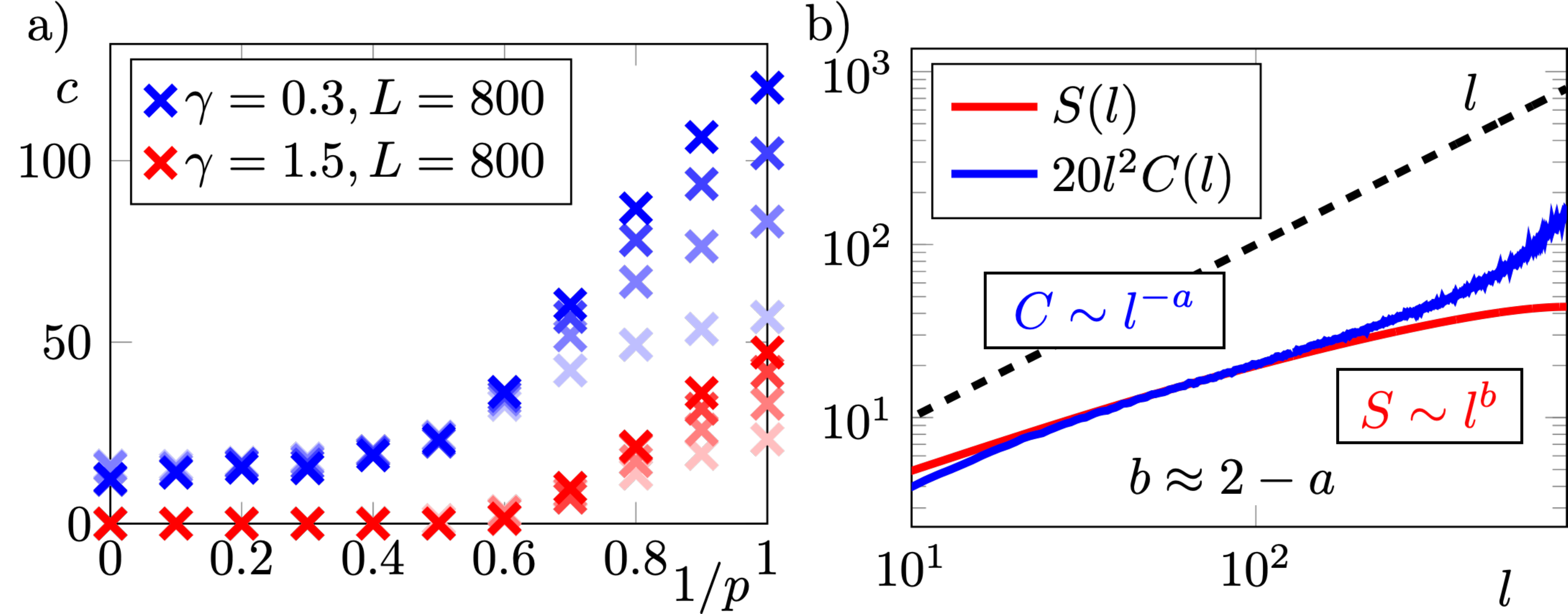}
  \caption{a) Effective central charge $c$ determined from a fit at $l \in [L/4,3L/4]$ for system sizes $L=200,400,600,800$. For $1/p \lesssim 0.6$, $c$ saturates, while it diverges with $L$ for $1 > 1/p \gtrsim 0.6$, indicating faster than $\log$ entanglement growth. b) Entanglement entropy and density-density correlation function deep in the algebraic scaling phase ($\gamma=0.3,p=1.25$) for $L=1600$. At intermediate distances, where finite-size effects are negligible, the algebraic scaling matches the analytical prediction $b= 2-a$.
  }
  \label{Fig3}
\end{figure}

Numerically, we distinguish the three phases, and their boundaries, by a combination of two approaches. (1) Informed by logarithmic growth of the entanglement entropy in the CFT phase, we assume a logarithmic scaling of $S(l,L)$ and extract a size-dependent effective central charge $c(L)$. In the area-law phase and in the CFT phase, $c(L)$ saturates as a function of the system size to $c(L)=0$ or $c(L)>0$. In the algebraic phase it does not saturate, indicating faster than logarithmic growth of $S(l,L)$ on large distances (see Fig.~\ref{Fig1}c, Fig.~\ref{Fig3}a), allowing us to localize the phase boundaries. (2) We make a general ansatz for $S(L/2,L)$ and $C(L/2,L)$, including both algebraic and logarithmic scaling, and then compute the best fit as a function of $L$ (see Fig.~\ref{Fig2}). This yields consistent results for the scaling behavior in the different regimes, as well as for the scaling exponents and the location of the phase boundaries (see Fig.~\ref{Fig1}d,e). In addition, we confirm these results via the behavior of $S(l,L)$ and $C(l,L)$ as functions of the subsystem size $l$ (see~\cite{supp}).

This existence of three distinct phases is also reflected in the form of $H_{\text{eff}}$. The Hamiltonian is composed of a free part, quadratic in $\hat\theta_x, \hat\phi_x$, and two competing nonlinearities. The structure of the dark state $|\psi_D\rangle$ then depends on whether or not the nonlinear contributions are relevant on large distances. Both nonlinearities tend to suppress fluctuations of their arguments, and to pin the corresponding fields. However, $\hat{\theta}_x, \hat\phi_x$ are conjugate and cannot be pinned simultaneously, i.e., at most one nonlinearity can be relevant. This gives rise to the three scenarios outlined above: (i) in the area law phase, $\lambda$ is relevant and pins the density field $\hat \phi_x$ to a measurement eigenstate, (ii) in the CFT phase, neither $\lambda$ nor $\Delta$ are relevant, yielding scale invariant free bosons, and (iii) in the algebraic scaling phase, the long-range hopping $\Delta$ is relevant and yields a pinning of the relative phases $\hat\theta_x-\hat\theta_{x+l}$ of the fermions. This is associated with long-range correlations in $\hat \theta_x$ and increased local particle number fluctuations $\sim\hat\phi_x$. 

The regimes (i) and (ii) have been established in the limit $p\rightarrow\infty$ in previous works (numerically in Refs.~\cite{alberton2021enttrans, bao2021symmetry} and analytically in Ref.~\cite{buchhold2021effective}). Apart from a $p$-dependent shift of the critical point towards larger values of $\gamma$ due to an enhanced kinetic energy, this picture remains unmodified for $3/2<p<\infty$. We thus focus on regime (iii), corresponding to an algebraic scaling of the entanglement entropy and the phase transitions (i)$\leftrightarrow$(iii) and (ii)$\leftrightarrow$(iii). Canonical power counting yields the scaling dimension $[\Delta]=3-2p$, indicating that the long-range aspect of the hopping is irrelevant for $p>3/2$, and confining the algebraic phase to $p<3/2$. The independence of the critical value $p_c=3/2$ of the measurement strength $\gamma>0$ -- implying that even frequent local measurements cannot overcome the entanglement generation of a long-ranged kinetic Hamiltonian -- is confirmed by the simulations, e.g., shown in Fig.~\ref{Fig1}d,e.

\textit{Characterizing the phase transitions.} -- 
The three different phases discussed above are separated from each other by three different transition lines, each corresponding to a different type of phase transition (illustrated in Fig.~\ref{Fig1}). Each phase transition  features a characteristic competition between different parts of the effective Hamiltonian: the quadratic part $(\sim\eta, \eta^{-1})$ tends to balance the fluctuations of $\hat\phi_x, \hat\theta_x$, while the nonlinearities $(\sim \Delta,\lambda)$ suppress fluctuations of $\hat\theta_x$ or $\hat\phi_x$. This competition is manifest in the perturbative renormalization group (RG) equations for $\eta, \Delta, \lambda$. 

The RG equations are obtained by rescaling spacetime with a factor $e^{-s}$, i.e., $dx,dt\rightarrow dx e^{-s},dt e^{-s} $, with $s$ infinitesimal, and then integrating out fast modes with momentum $\Lambda e^{-s}<|q|<\Lambda$ (here $\Lambda=\pi$ is the short distance cutoff). Up to first order in the nonlinear couplings, this yields (see~\cite{supp})
\begin{align}
   \partial_s\Delta&=(3-2p-\eta)\Delta,\label{eq:RGlr}\\
    \partial_s\eta&=-\eta^2\Delta,\label{eq:RGquad}
\end{align}
A similar set of RG equations has been obtained in Ref.~\cite{maghrebi2017} for the ground state of a Hermitian, long-range interacting XXZ-chain. In that case, however, all couplings are real, and the canonical scaling dimension of the long-range coupling $\Delta$ is modified compared to Eq.~\eqref{eq:RGlr} by replacing $p\rightarrow p/2$. From this perspective, the measurement-induced quantum phase transitions observed here represent a generalization of ground state phase transitions of interacting particles, to dark state phase transitions in non-unitary, measurement-induced dynamics. 

The RG equations \eqref{eq:RGlr},\eqref{eq:RGquad} yield several important insights: (a) For $p>3/2$, the long-range hopping $\sim\Delta$ is always irrelevant and any initial $\Delta\neq0$ decays to zero exponentially fast ($\text{Re}(\eta)\ge0$ is required for stability). In this case, the monitored long-range hopping dynamics are effectively reduced to a nearest neighbor hopping model, such as discussed in Refs.~\cite{alberton2021enttrans, buchhold2021effective}. It therefore undergoes a Berezinskii-Kosterlitz-Thouless (BKT) transition from a critical to an area law phase as a function of the measurement strength. (b) For $p\le 3/2$, the long-range hopping is relevant and strongly enhances fluctuations of $\phi_x$, i.e., rapidly decreases $\eta$ in Eq.~\eqref{eq:RGquad}, already on the level of first order RG equations~\footnote{For a conventional sine-Gordon model with short-range nonlinearity, the renormalization of $\eta$ starts at the level of second order perturbation theory.}. This confirms the result from canonical power counting. (c) Although Eq.~\eqref{eq:RGlr} is reminiscent of a conventional sine-Gordon model (here for a $p$-dependent canonical scaling dimension), the transition between the critical and the long-range phase does not fall into the BKT paradigm. This roots in the linear appearance of $\Delta$ in Eq.~\eqref{eq:RGquad}, obtained at first order perturbation theory. It gives rise to an unconventional transition, which is characteristic for long-range coupled systems~\cite{maghrebi2017}. (d) A similar argument applies for the transition from the area law to the long-range phase. This transition arises from the direct competition between the two cosine terms. For a short-ranged sine-Gordon model, this gives rise to a sequence of universality classes depending on the factor in the nonlinearities, including the Ising and parafermionic ones \cite{Lecheminant}. However, due to Eq.~\eqref{eq:RGquad}, this is again different for the long-range model, where a transition of this type has not yet been characterized. 

\textit{Characterizing the algebraic scaling phase.} -- For $p>3/2$, the long-range hopping, and the coupling $\Delta$, are relevant. The dark state $|\psi_D\rangle$, as well as the fermion entanglement entropy and correlation functions, are then modified by the impact of $\hat{H}_{\text{lr}}$ in Eq.~\eqref{eq:ReplicaHamiltonian}. It pushes the phases $\hat\theta_l$ of the fermion operators $\hat{c}_l, \hat{c}^\dagger_l$ to align with each other, and to pin the argument of $\cos \sqrt{2}(\hat\theta_x-\hat\theta_y)$ at small values. Expanding $\hat{H}_{\text{eff}}$ up to second order in $\hat\theta_x$ then yields (in Fourier space)
\begin{equation}  \label{eq:Hlrscaling}
    \hat{H}_{\text{eff}}=\int_qq^2 \left\{-i\Delta_p|q|^{2p-3}\hat\theta_{-q}\hat\theta_q+\eta\hat\phi_{-q}\hat\phi_q \right\},
\end{equation}
with a positive integration constant $\Delta_p$ (see~\cite{supp}). The terms $\sim\hat\theta_{q}$ in $\hat{H}_{\text{eff}}$ become more and more dominant as $q\rightarrow0$. This then also governs the dark state $|\psi_D\rangle$ at long wavelength (see~\cite{supp}), such that the correlation functions and the entanglement entropy acquire a $p$-dependent scaling exponent,
\begin{align}
    S(l)&\sim B |l|^{b}+s_0,\quad b = \tfrac{3}{2}-p, \label{eq:SscalingAn}\\
    C(l)&\sim A |l|^{-a}, \qquad\, a= p+\tfrac{1}{2}. \label{eq:CscalingAn}
\end{align}
This analytical estimate is confirmed by the numerical simulations very accurately, which is demonstrated in Figs.~\ref{Fig1}d,e, Fig.~\ref{Fig2} (purple lines), and~\cite{supp}. These scaling exponents clearly differ from the volume-law $b=1$, found for instance in monitored random circuits~\cite{gullans2019,Zabalo2020,zhang2020,Tang2020,Bao_2020,chen2020,Gullans2020,Nahum20a}. We find a slower growth of the entanglement entropy $b<1/2$, matching the scaling relation $b= 2-a$ (see Fig.~\ref{Fig3}b), implied by Eq.~\eqref{eq:scanalyyt}.

{\it Conclusion.--} Long-range dynamics leads to a novel measurement induced sub-volume phase with a maximum entanglement growth exponent $S \sim \sqrt{L}$. Despite the enhanced entangling effect of long range hoppings, the system does not reach a volume law characteristic of an ergodic phase. 
Indeed, a volume law is typically associated with excited or generic states, located in the middle of the spectrum of a Hamiltonian. Here rather, the trajectory wave functions evolve to states that bear strong similarities to ground states. In fact, it is intriguing to notice the similarity of the phase structure to the ground state phase diagram of long-range interacting spin models in one dimension \cite{maghrebi2017}, where the density pinning effect of measurements is replaced by nearest neighbor interactions. This suggests a persistence of quantum ground state dynamics in the trajectory wave function, which is reflected in the effective {\it cooling} towards measurement-induced dark states in the replica formalism. It will be exciting to identify the precise criteria for the occurrence of true volume law phases or classical vs. quantum phase transitions in trajectory ensembles, e.g. in terms of the Gaussianity of the state present here, or of the integrability of the generator of dynamics \cite{ziolkowska2020,piroli2020,lunt2020}.

{\it Note added:} During the completion of this work, we became aware of two complementary works on the measurement-induced transition in long-range coupled Hamiltonian systems~\cite{minato2021fate,block2021measurementinduced}, where a critical hopping exponent $p_c$ relatable to our $p_c=3/2$ was identified.

\begin{acknowledgments}
  \section*{Acknowledgements}
  We acknowledge support from the Deutsche Forschungsgemeinschaft (DFG, German Research Foundation) under Germany's Excellence Strategy Cluster of Excellence Matter and Light for Quantum Computing (ML4Q) EXC 2004/1 390534769, and by the DFG Collaborative Research Center (CRC) 183 Project No. 277101999 - project B02. Furthermore we acknowledge support by the European Research Council (ERC) under the Horizon 2020 research and innovation program, Grant Agreement No. 647434 (DOQS). M.B. acknowledges funding via grant DI 1745/2-1 under DFG SPP 1929 GiRyd. The code for our numerical computations was implemented in Julia~\cite{bezanson17}. We thank A.~Chiocchetta, M.~Gullans, and D.~Huse for fruitful discussions.
\end{acknowledgments}

\bibliography{EntEnt}
\end{document}


\date{\today}

\title{Supplemental material for "Measurement-induced dark state phase transitions in long-ranged fermion systems"}

\author{T. Müller}
\affiliation{Institut f\"ur Theoretische Physik, Universit\"at zu K\"oln, D-50937 Cologne, Germany}
\author{S. Diehl}
\affiliation{Institut f\"ur Theoretische Physik, Universit\"at zu K\"oln, D-50937 Cologne, Germany}
\author{M. Buchhold}
\affiliation{Institut f\"ur Theoretische Physik, Universit\"at zu K\"oln, D-50937 Cologne, Germany}

\begin{abstract}
	We provide the following additional information: (i) A derivation of the effective non-hermitian long-range hopping Hamiltonian in the relative replica representation in leading order perturbation theory. (ii) A discussion of how the fermion observables of the underlying model can be related to analytically accessible quantities in the bosonized language. (iii) Details on the numerical procedure to extract scaling exponents, phases and phase boundaries. (iv) A discussion of the infrared scaling of the long-range hopping Hamiltonian with the appearance of the exact position of the transition into the algebraic scaling phase, $p=3/2$. (v) Evaluation of the boson correlation functions in the dark state of the non-hermitian Hamiltonian. (vi) Derivation of the RG flow due to the long-range hopping.
\end{abstract}

\maketitle

\section{Replica Hamiltonian for long-range hopping} \label{app:LongRangeHamiltonian}

Here, we sketch the derivation of the long-range part of the replica Hamiltonian. We begin with the bosonized Hamiltonian that acts on individual replicas $ \hat{H}^{(i)} \simeq \hat{H}^{(i)}_\text{sr}  +  \hat{H}^{(i)}_\text{lr}$, containing a local, quadratic part due to the short-ranged part of the hopping $\hat{H}^{(i)}_\text{sr}$, and a non-linearity due to the long-range part of the hopping (we set the lattice spacing $a=1$)
\begin{subequations}
\begin{align}
    \hat{H}_\text{lr}^{(i)} = - \delta \int_{x,\vert y \vert >1} \vert y \vert^{-p} \cos \left( \hat{\theta}_{x+y}^{(i)}-\hat{\theta}_{x}^{(i)} \right) ,
\end{align}
\end{subequations}
with $\delta>0$ quantifying the strength of the long-range term. The corresponding Hamiltonian acting on two replicas is $\hat{H}^{(R_2)}_\text{lr}= \hat{H}^{(1)}_\text{lr}+\hat{H}^{(2)}_\text{lr}$. Applying the rotation in the replica basis $\hat{\phi}^{(a,r)}_x=\tfrac{1}{\sqrt{2}}(\hat{\phi}^{(1)}_x\pm \hat{\phi}^{(2)}_x),\ \ \hat{\theta}^{(a,r)}_x=\tfrac{1}{\sqrt{2}}(\hat{\theta}^{(1)}_x\pm\hat{\theta}^{(2)}_x)$ reveals that the non-linearity couples the center-of-mass coordinate $\hat{\theta}^{(a)}$ and the relative coordinate $\hat{\theta}^{(r)}$
\begin{equation}
    \hat{H}^{(R_2)}_\text{lr} = - 2 \delta \int_{x, \vert y \vert >1} \vert y \vert^{-p} \cos \left( \frac{\hat{\theta}_{x+y}^{(a)}-\hat{\theta}_{x}^{(a)}}{\sqrt{2}} \right) \cos \left( \frac{\hat{\theta}_{x+y}^{(r)}-\hat{\theta}_{x}^{(r)}}{\sqrt{2}} \right).
\end{equation}
Translating this term into a Keldysh field theory description~\cite{Sieberer_2016} yields
\begin{multline}
    S_\text{lr}^{(R_2)} =2 \delta \int_{t,x,\vert y \vert >1} \Bigg[ \cos \left( \frac{\theta_{+,x+y}^{(a)}-\theta_{+,x}^{(a)}}{\sqrt{2}} \right) \cos \left( \frac{\theta_{+,x+y}^{(r)}-\theta_{+,x}^{(r)}}{\sqrt{2}} \right) - \\ - \cos \left( \frac{\theta_{-,x+y}^{(a)}-\theta_{-,x}^{(a)}}{\sqrt{2}} \right) \cos \left( \frac{\theta_{-,x+y}^{(r)}-\theta_{-,x}^{(r)}}{\sqrt{2}} \right) \Bigg] \times \vert y \vert^{-p},
\end{multline}
with $\pm$ indicating forward and backward time-evolution on the Keldysh contour. Next, we integrate out the center-of-mass fields $\theta^{(a)}_\pm$ perturbatively in $\delta$. To first order, we need to evaluate $\left\langle S_\text{lr}^{(R_2)} \right\rangle_{(a)}$. This term vanishes because
\begin{equation}
    \left\langle \cos \left( \frac{\theta_{\pm,x+y}^{(a)}-\theta_{\pm,x}^{(a)}}{\sqrt{2}} \right) \right\rangle_{(a)} = \exp{-\frac{1}{4} \left\langle \left( \theta_{\pm,x+y}^{(a)}-\theta_{\pm,x}^{(a)} \right)^2 \right\rangle_{(a)}} = 0,
\end{equation}
due to the heating of the absolute component to an infinite temperature state due to the monitoring (cf. Ref.~\cite{buchhold2021effective}). For that reason, the first non-vanishing contribution to the effective relative replica action $ S_\text{lr}^{(r)}$ appears only at second order in $\delta$,  and renders
\begin{equation}
   2i \delta^2 \int_{t,x,\vert y \vert > 1} \Bigg[ \cos^2 \left( \frac{\theta^{(r)}_{+,x+y}-\theta^{(r)}_{+,x}}{\sqrt{2}} \right) + \cos^2 \left( \frac{\theta^{(r)}_{-,x+y}-\theta^{(r)}_{-,x}}{\sqrt{2}} \right) \Bigg] \times \vert y \vert^{-2p}.
\end{equation}
The doubling of the exponent $p$ results due to its origin in second order perturbation. With the identification $S=-\int_t (H^*_+ - H_-)$, and $\Delta \equiv \delta^2$, we deduce the effective Hamiltonian for the relative coordinate in the replica-basis, presented in Eq.~(3c) of the main text.

\section{Fermion observables}\label{app:FermiObs}
We compare the numerical results for the fermion observables $C(l), S(l)$ with the analytical predictions obtained from the dark state $|\psi_D\rangle$ of $H_{\text{eff}}$ in Eq.~(3a) in the thermodynamic limit $L\rightarrow\infty$ (in this case the argument $L$ is dropped). The fermion density-density correlation function $C(l)$ is then obtained via the bosonization identity $n_x\rightarrow-\frac{\partial_x\phi_x}{\pi}+\text{h.h.}$, where h.h. indicates contributions from higher harmonics. These contributions are then neglected. This yields $ C(l)=\tfrac{1}{2\pi^2}\partial_l^2\langle\phi_x\phi_{x+l}\rangle$. Computing the fermion entanglement entropy analytically is more subtle. For Dirac fermions in a Gaussian ground state (corresponding to a free theory with compactification radius $K=1$), it was shown in Refs.~\cite{Casini_2009, Calabrese_2004} that the subsystem entanglement entropy can be computed from the boson correlation function, i.e., 
\begin{equation}
    S(l)\overset{\text{free Dirac}}{=}\frac{1}{3}\langle\phi_x\phi_{x+l}\rangle,\label{eq:EntApprox}
\end{equation}
 yielding Eq.~(5) in the main text. During the monitoring, each individual wave function corresponds to a Gaussian state and therefore, by assuming Dirac fermions, we approximate the entanglement entropy in each phase by Eq.~\eqref{eq:EntApprox}. This yields a remarkably good agreement between the numerically obtained fermion entanglement and the boson theory. The relation~\eqref{eq:EntApprox} between the boson correlation functions and the fermion entanglement entropy has also been highlighted in Refs.~\cite{bao2021symmetry, jian2021syk}, where both sides of the equation describe the free energy of a pair of (half-) vortices in the respective formalism.

\section{Numerical evaluation of phase transition locations and scaling exponents} \label{app:NumericsSupplement}

\begin{figure}
  \includegraphics[width=1\linewidth]{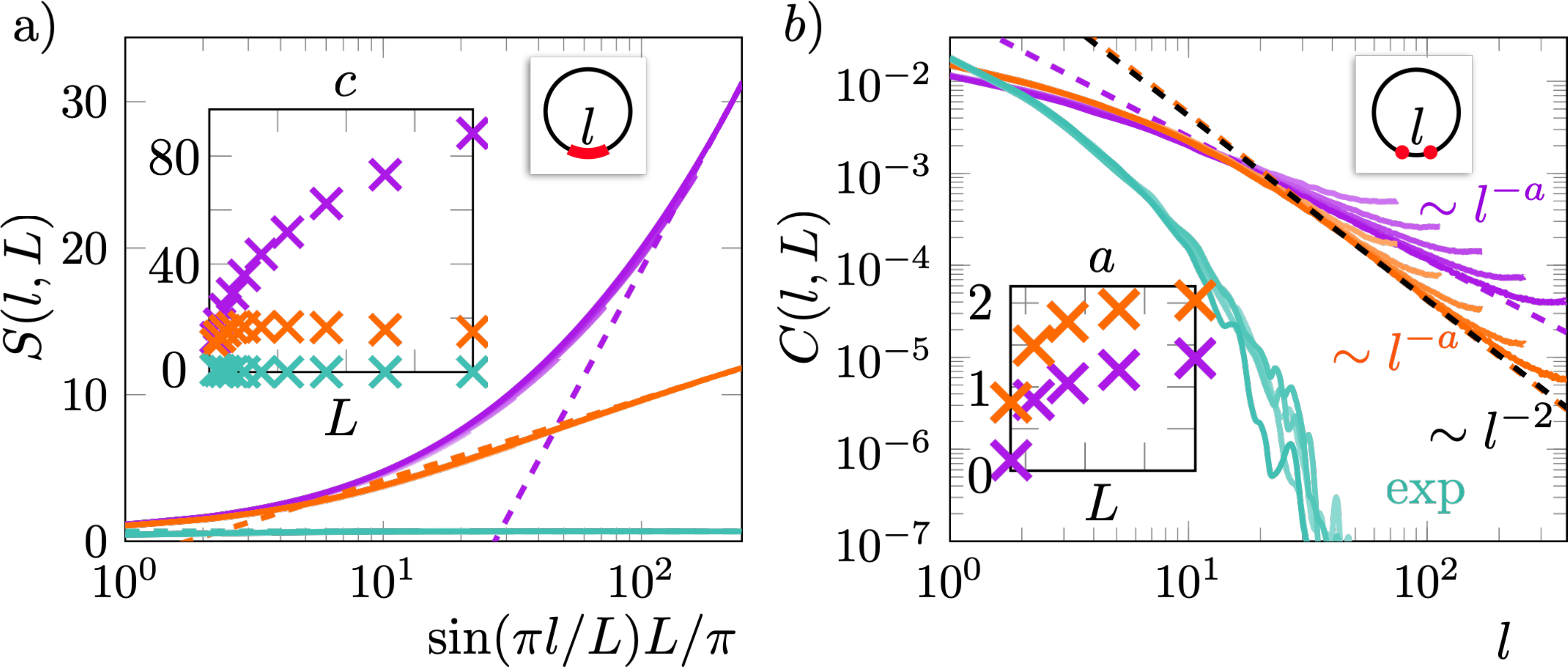}
  \caption{Further characterization of CFT ($\gamma=0.3, p=1.25$, orange), area-law ($\gamma=2,p=5$, light blue) and algebraic scaling phase ($\gamma=0.3,p=5$, purple) in terms of subsystem size dependent observables for different system sizes (different shades) up to $L=770$. a) The collapse of the entanglement entropy onto a function of the conformally invariant combination $\frac{L}{\pi} \sin \frac{\pi l}{L}$ for different $L$ breaks down in the algebraic scaling phase. Inset: $L$-dependence of the extracted effective central charge. Dashed lines: best fit under the assumption of a CFT at $L=770$. b) Emergence of algebraic scaling of the correlation function at intermediate $l$ for $L \rightarrow \infty$ in the CFT and the algebraic scaling phase. Fitting to the scaling regime yields the exponents indicated in the inset and (for $L=770$) by dashed lines in the main plot (black dashed line $\sim l^{-2}$ for comparison).
  }
  \label{FigApp1}
\end{figure}

First, we recall the properties of phases appearing in the short-range hopping limit~\cite{alberton2021enttrans}. In the CFT-like regime (orange curves in Fig.~\ref{FigApp1} and Fig.~2 of the main text), we observe an asymptotic scaling collapse of the entanglement entropy,
\begin{equation}  \label{Appeq:logscaling}
    S(l,L) \simeq \frac{c}{3}  \log \left[ \frac{L}{\pi} \sin{\frac{\pi l}{L}} \right] + s_0,
\end{equation}
familiar from a CFT with periodic boundary conditions in 1+1 dimensions~\cite{Calabrese_2004,Calabrese_2009}. The effective central charge $c$ can be extracted efficiently from the simulations (see Fig.~\ref{FigApp1}a, inset) by fitting the data to Eq.~\eqref{Appeq:logscaling} for $L/4<l<3 L/4$, and depends in a continuous way on both the hopping range $p$ and the monitoring strength $\gamma$ (see Fig.~1c in the main text). However, the phase is stable against deviations from the nearest-neighbor hopping $p=\infty$. This result is supported by the asymptotic scaling collapse of the correlation function onto
\begin{equation} \label{Appeq:CFTScaling}
    C(l,L) \sim \left[ \frac{L}{\pi} \sin{\frac{\pi l}{L}} \right]^{-2} \sim \frac{1}{l^2} , \frac{1}{L^2},
\end{equation}
in the same parameter-regime (see Fig.~\ref{FigApp1}b) and Fig.~2b in the main text), in agreement with conformal scaling~\cite{Calabrese_2004,Calabrese_2009}. 

Conversely, the area-law phase (light blue lines in Fig.~\ref{FigApp1} and Fig.~2 in the main text) is characterized by asymptotically constant entanglement entropy, quantified by a vanishing effective central charge, and exponentially decaying correlation functions, both as functions of $l$ and $L$.

Applying the same procedure in the long-range hopping regime (purple curves in Fig.~\ref{FigApp1} and Fig.~2 in the main text) reveals a breaking of this behavior in three ways: (i) $S(l,L)$ and $C(l,L)$ do not collapse onto a function of the scale-invariant combination $\frac{L}{\pi} \sin{\frac{\pi l}{L}}$, (ii) the extracted effective central charge does not approach a finite value for for $L \to \infty$ (see Fig.~3a in the main text), indicating algebraic scaling of the entanglement entropy, and (iii) the correlation function $C(l,L)$ decays algebraically, but slower than $l^{-2}$ or $L^{-2}$. To extract the critical point, where the algebraic scaling sets in, and the exponents for both $S$ and $C$ in this regime, we use an ansatz to capture both conformal scaling and algebraic scaling, and finite size effects
\begin{equation}
    S(L/2,L) = B L^b + \frac{c}{3} \log \frac{L}{\pi} + s_0, \quad  C(L/2,L) = \frac{1}{A L^a + D L}.
\end{equation}
Fitting $B,b,c,s_0,A,a$ and $D$ to the numerical data is sensitive to the phase transition. Especially the exponents $a$ and $b$ can be extracted quantitatively, signalling the algebraic scaling phase by $a<2$ and $b>0$ (cf. Fig.~1d,e in the main text). Comparing the exponents extracted in this way to $S(l,L) \sim l^b$ and $C(l,L) \sim l^{-a}$ in a scaling regime at intermediate $l$ shows good agreement (cf. Fig.~\ref{FigApp1}).

\section{Scaling form of long-range Hamiltonian} \label{App:ScalingHamiltonian}

If the long-range hopping Hamiltonian shown in Eq.~(3c) of the main text is relevant in the RG sense, we may assume a large parameter $\Delta$ and hence a pinning of $\hat{\theta}_x$ to a constant. Expanding around this constant (that we set to $0$ for convenience) yields to leading order and up to an additive constant
\begin{equation}
    \hat{H}_\text{lr} \simeq -i\Delta \int_{x, \vert y \vert>1} \vert y \vert^{-2p} \left( \hat{\theta}_x-\hat{\theta}_y \right)^2.
\end{equation}
Applying the Fourier transformation yields
\begin{equation}
    \hat{H}_\text{lr} \simeq - 8 i \Delta \int_q \hat{\theta}_{-q} \hat{\theta}_q \int_1^\infty dy \frac{\sin^2 \frac{qy}{2}}{y^{2p}} .
\end{equation}
The canonical scaling dimension is then extracted from the integral
\begin{equation} \label{eq:Integral}
    \int_1^\infty dy \frac{\sin^2 \frac{qy}{2}}{y^{2p}} = \vert q \vert^{2p-1} \int_{q}^\infty ds \frac{\sin^2 \frac{s}{2}}{s^{2p}}.
\end{equation}
For $1<p<3/2$, taking limit $q \to 0$ yields a convergent dimensionless integral $I_p \equiv \int_0^\infty ds \sin^2 (s/2) s^{-2p}>0$, such that the momentum dependence is entirely in the prefactor of the integral, and we find
\begin{equation}
    \hat{H}_\text{lr} \simeq -i \Delta_p \int_q \vert q \vert^{2p-1} \hat{\theta}_{-q} \hat{\theta}_q,
\end{equation}
with $\Delta_p = 8 \Delta I_p > 0$. Since $2p-1<2$, this is more relevant than the term $\sim q^2 \hat{\theta}_{-q} \hat{\theta}_q$ from the short-range Hamiltonian (Eq.~(3a) in the main text), such that we drop the latter. The fact that $\hat{H}_\text{lr}$ and the non-linearity in $\hat{H}_\text{sr}$ do not commute implies that only one of the two (in this case $\hat{H}_\text{lr}$) can be relevant. Together, we find the effective Hamiltonian~(8) in the main text. 

Conversely, if $p>3/2$, the integral $I_p$ is divergent and has to be regularized by re-inserting $q$ as a cutoff. The divergence of the integral precisely cancels the $p$-dependence of the prefactor, and we obtain $\sim q^2 \hat{\theta}_{-q} \hat{\theta}_q$ from the long-range term, which renormalizes the parameters of the short-range Hamiltonian.

\section{Correlation functions} \label{App:Correlation functions}

Here, we evaluate the $\hat{\phi} \hat{\phi}$ correlation function in the dark state $|\psi_D\rangle$ of the linearized replica Hamiltonian in the long-range regime
\begin{equation}
    \hat{H}_\text{eff} = \frac{1}{2\pi} \int_q \left\lbrace -i \Delta_p \vert q \vert^{2p-1} \hat{\theta}_{-q} \hat{\theta}_q + \eta q^2 \hat{\phi}_{-q} \hat{\phi}_q \right\rbrace,
\end{equation}
where we assume $1<p<3/2$. We represent this in terms of bosonic operators $\hat{b}_q,\hat{b}_q^\dagger$ with $[\hat{b}_p,\hat{b}_q^\dagger]=\delta_{p,q}$ and find
\begin{subequations}
\begin{align}
    \hat{H}_\text{eff} &= \int_q \left( \begin{array}{cc}
        \hat{b}_q^\dagger & \hat{b}_{-q} 
    \end{array} \right) M_q \left( \begin{array}{c}
         \hat{b}_q  \\
         \hat{b}_{-q}^\dagger 
    \end{array} \right), \\
    M_q &= \frac{\vert q \vert}{4} \left( \begin{array}{cc}
        \eta+i \Delta_p \vert q \vert^{2p-3} & \eta- i \Delta_p \vert q \vert^{2p-3} \\
        \eta-i \Delta_p \vert q \vert^{2p-3}  & \eta+i \Delta_p \vert q \vert^{2p-3}
    \end{array} \right).
\end{align}
\end{subequations}
By a Bogoliubov transformation
\begin{equation}
    \left( \begin{array}{c}
         \hat{b}_q  \\
         \hat{b}^\dagger_{-q}
    \end{array} \right) = \left( \begin{array}{cc}
        \alpha^{*}_{-q} & -\beta_q \\
        -\beta^{*}_{-q} & \alpha_{q}
    \end{array} \right) \left( \begin{array}{c}
         \hat{c}_q  \\
         \hat{c}^{\dagger}_{-q}
    \end{array} \right),
\end{equation}
with $\vert \alpha_q \vert^2-\vert \beta_q \vert^2=1$, we bring the Hamiltonian into a tri-diagonal form in terms of the bosonic operators $\hat{c}_q,\hat{c}_q^{ \dagger}$,
\begin{equation}
    \hat{H} = \epsilon_q \left( \hat{c}_q^\dagger \hat{c}_q + \hat{c}_{-q} \hat{c}_{-q}^\dagger \right) + \eta_q \hat{c}_q \hat{c}_{-q},
\end{equation}
with $\Im \epsilon_q<0$, such that the excitations in the basis of $\hat{c}_q,\hat{c}_q^\dagger$ decay exponentially in time. The term proportional to $\eta_q$ ensures that the system cannot be frozen in a different state. For that reason, we identify $\ket{\psi_D}$, defined by $\hat{c}_q \ket{\psi_D}=0$ as the dark state of the non-Hermitian Hamiltonian in the relative replica coordinate, and hence the state that determines the correlation functions in the stationary limit $t \rightarrow \infty$. The tridiagonal form demands 
\begin{equation}
    \alpha_q = - \beta_q = \frac{1}{2} \left( \frac{i \Delta_p}{\eta} \right)^{1/4} \vert q \vert^{p/2-3/4} + \mathcal{O}(q^0).
\end{equation}
Higher orders in $q$ ensure the validity of the Bogoliubov transformation. In terms of the Luttinger liquid operators, we find
\begin{align}
    \bra{\psi_D} \hat{\phi}_{-q} \hat{\phi}_q \ket{\psi_D} & = \frac{\pi}{2 \vert q \vert} \left[ 1+ 2 \vert \beta_q \vert^2 - \alpha_q \beta_q^{*} - \beta_q \alpha_q^{*} \right] \notag \\
    &=  \frac{3\pi}{2} \sqrt{ \frac{\Delta_p}{\vert \eta \vert}} \vert q \vert^{p-5/2} + \mathcal{O}(\vert q \vert^{-1}).
\end{align}
By a Fourier-transformation, we find the scaling $\langle \hat{\phi}_{x+l} \hat{\phi}_x \rangle \sim l^{3/2-p}$ from the main text (Eqs.~(9),(10)).

\section{First order renormalization group equations} \label{App:RG}
We briefly review the first order perturbative renormalization group (RG) approach for sine-Gordon type models to motivate the form of the RG equations Eqs.~(6)-(8) in the main text. We display the steps for the first order correction to the Hamiltonian induced by $\hat H_{\text{lr}}$. Canonical power counting for the terms in this Hamiltonian yields $[\Delta]=3-2p$ ($3$ counting two space integrals, one time integral). The first order RG correction is obtained from decomposing the fields $\hat \theta_x= \hat \theta_x^<+\hat \theta_x^>$ into fast, short distance modes $\hat \theta_x^<=\int_{\Lambda e^{-s}<|k|<\Lambda}e^{ikx} \hat \theta_x$ and slow, long distance modes $\hat \theta_x^>=\int_{|k|<\Lambda e^{-s}}e^{ikx} \hat \theta_x$. Here $\Lambda=\pi/a$ is the short-distance cutoff with the lattice spacing $a=1$.

The first order correction to the Hamiltonian $\Delta \hat H_{\text{lr}}$ is then obtained by taking the average with respect to short distance modes, assuming they are in the dark state of the quadratic part of $H_{\text{eff}}$. The cosine renormalizes multiplicatively 
\begin{align}
  &\langle \cos\left[\sqrt{2}(\hat\theta_x-\hat\theta_y)\right]\rangle_<=\cos\left[\sqrt{2}(\hat\theta_x^>-\hat\theta_y^>)\right] e^{-\langle (\hat \theta_x^<- \hat \theta_y^<)^2\rangle_<}\nonumber\\
  &=\cos\left[\sqrt{2}(\hat\theta_x^>-\hat\theta_y^>)\right] e^{-2\langle (\hat \theta_x^<)^2\rangle_<}(\underbrace{1}_{(\text{i})}+\underbrace{e^{2\langle \hat \theta_x^< \hat \theta_y^<\rangle_<}-1}_{(\text{ii})}).\label{eq:RGEQ}
\end{align}
Here $\langle ...\rangle_<$ denotes the average with respect to the dark state $|\psi_D^<\rangle=\otimes_{\Lambda e^{-s}<|k|<\Lambda}|\psi_D^k\rangle$, where $|\psi_D^k\rangle$ is the $k$-momentum dark state of the quadratic part of $H_{\text{eff}}$. The first term (i) in Eq.~\eqref{eq:RGEQ} then represents the conventional renormalization of a $\cos$-nonlinearity, which one would obtain also for purely local terms. The second term (ii), however, is characteristic for the long-range Hamiltonian with two-different arguments $x,y$. Since both $\hat \theta_x^<, \hat \theta_y^<$ are short distance modes, the average is only non-zero if $x\approx y$. We can therefore expand $y$ around $x$, and we approximate the short distance terms by their leading order contribution $y\approx x+a$. Besides being linear in $\Delta$, this term is then of the same form as the common {\it second order} perturbative correction in the conventional, local sine-Gordon model. This yields
\begin{align}
  \langle \cos\left[\sqrt{2}(\hat\theta_x-\hat\theta_y)\right]\rangle_<\approx& \Big\{e^{-\eta s}\cos\left[\sqrt{2}(\hat\theta_x^>-\hat\theta_y^>)\right]\nonumber\\ &-\delta(|y-x|-a)(a\partial_x \hat \theta_x^>)^2F_a\Big\}.
\end{align}
where $F_a=e^{-\eta s}(e^{2\langle \hat \theta_x^< \hat \theta_{x+a}^<\rangle_<}-1)$. Together with the canonical power counting this yields the perturbative flow equations 
\begin{align}
    \partial_s \Delta&=(3-2p-\eta)\Delta,\\
    \partial_s\eta^{-1}&= \Delta F_a.
\end{align}
Assuming a negligible dependence of $F_a$ on $s$ and rescaling $\Delta\rightarrow\Delta F_a$ yields the RG equations~(6),(7) from the main text. 

\bibliography{EntEnt}